\documentclass[journal]{IEEEtran}

\usepackage[]{graphicx}
\usepackage{caption}
\usepackage{subfig} %should come after caption package if given, can't be used with caption2, defines subfloat
\usepackage{amsmath}
\usepackage{amssymb}
\usepackage{epsfig}
\usepackage{cite}
\usepackage{color}
\usepackage{balance}
\usepackage{amsmath}
\usepackage{amsfonts} % for the \checkmark command 
\usepackage{graphicx}
\usepackage{pdfpages}
\usepackage[T1]{fontenc} % For less than and greater than sign
\usepackage{array}
\usepackage{url}

\usepackage{color}
\usepackage{wrapfig}
\usepackage{lipsum}
\usepackage[hidelinks]{hyperref}
\usepackage{multirow}
\usepackage{tabularx}
\usepackage{tikz}

\usepackage{algpseudocode} % For algorithm 
\usepackage{algorithm,algpseudocode}

\begin{document}
\title{CloudSegNet: A Deep Network for \\Nychthemeron Cloud Image Segmentation}

\author{
Soumyabrata~Dev,~\IEEEmembership{Member,~IEEE,}
Atul~Nautiyal,\\
Yee~Hui~Lee,~\IEEEmembership{Senior~Member,~IEEE,} and~Stefan~Winkler,~\IEEEmembership{Fellow,~IEEE}% <-this % stops a space
\thanks{
Manuscript received November 08, 2018; revised March 03, 2019; and accepted March 24, 2019.
}

\thanks{
S.\ Dev and A.\ Nautiyal are with the ADAPT SFI Research Centre, Trinity College Dublin, Ireland (e-mail: soumyabrata.dev@adaptcentre.ie, nautiyaa@tcd.ie). Y.\ H.\ Lee is with the School of Electrical and Electronic Engineering, Nanyang Technological University, Singapore (e-mail: EYHLee@ntu.edu.sg). S.\ Winkler is with AI Singapore and the School of Computing, National University of Singapore (e-mail: winkler@comp.nus.edu.sg).
}% <-this % stops a space
%\thanks{
%Y. H. Lee is with School of Electrical and Electronic Engineering, Nanyang Technological University, Singapore (e-mail: EYHLee@ntu.edu.sg)
%}% <-this % stops a space
%\thanks{
%S. Winkler is with the Advanced Digital Sciences Center (ADSC), University of Illinois at Urbana-Champaign, Singapore (e-mail: Stefan.Winkler@adsc-create.edu.sg)
%}% <-this % stops a space
\thanks{
Send correspondence to Y.\ H.\ Lee, E-mail: EYHLee@ntu.edu.sg.}% <-this % stops a space
%\thanks{
%The ADAPT Centre for Digital Content Technology is funded under the SFI Research Centres Programme (Grant 13/RC/2106) and is co-funded under the European Regional Development Fund.
%}
}

\markboth{IEEE Geoscience and Remote Sensing Letters,~Vol.~XX, No.~XX, XX~2019}%
{Shell \MakeLowercase{\textit{et al.}}: Bare Demo of IEEEtran.cls for IEEE Journals}



\maketitle

\begin{abstract}
We analyze clouds in the earth's atmosphere using ground-based sky cameras. An accurate segmentation of clouds in the captured sky/cloud image is difficult, owing to the fuzzy boundaries of  clouds. Several techniques have been proposed that use color as the discriminatory feature for cloud detection. In the existing literature, however, analysis of daytime and nighttime images is considered separately, mainly because of differences in image characteristics and applications. In this paper, we propose a light-weight deep-learning architecture called \emph{CloudSegNet}. It is the first that integrates daytime and nighttime (also known as \emph{nychthemeron}) image segmentation in a single framework, and achieves state-of-the-art results on public databases.
\end{abstract}


% \fi
\IEEEpeerreviewmaketitle


% Included in addition to the bare file

%\title{CloudSegNet: A Deep Network for Nychthemeron Cloud Segmentation}

%
% Single address.
% ---------------
%\name{Author(s) Name(s)\thanks{Thanks to XYZ agency for funding.}}
%\address{Author Affiliation(s)}

%\begin{keywords}
%Cloud segmentation, whole sky imager, nychthemeron, deep learning.
%\end{keywords}

\section{Introduction}
\label{sec:intro}
\IEEEPARstart{C}{louds} play an important role for understanding the hydrological balance of nature and various events in the earth's atmosphere. Such studies are done mainly using satellite images, which generally suffer from low temporal and/or spatial resolution, however. 
Recently, with increasing demands from applications such as solar energy generation, high-resolution ground-based sky cameras are increasingly used, in addition to satellite and hyperspectral images. These ground-based cameras, popularly known as Whole Sky Imagers (WSIs), are able to collect much more frequent and more localized information about the clouds. Images captured by sky cameras provide a plethora of information to remote sensing analysts. Subsequently, a number of popular machine learning techniques \cite{dev2016ground} can be used on these images to understand different phenomena of cloud dynamics in the atmosphere. 
Likewise, deep learning techniques are now extensively used in remote sensing for this purpose, e.g.\ for scene classification \cite{cheng2018when} or geospatial object detection \cite{li2018deep}. 

However, detecting clouds in the images is  challenging because of the non-rigid structure of the cloud mass. As such, image segmentation techniques involving shape prior information are not applicable for this task. Existing techniques in the literature use color as the discriminatory feature for cloud detection \cite{li2011a,souza2006a,dev2017rough,long2006retrieving}. 

Most of the related works in cloud segmentation work on daytime images and use a combination of red and blue color channels. We have previously conducted a thorough analysis of color spaces and components used in cloud detection \cite{dev2014systematic}. 
A few techniques for nighttime sky/cloud image segmentation techniques have also been developed \cite{yang2009image,yang2010automatic}. However, none of them have been designed for or tested on sky/cloud images taken during both daytime and nighttime. This poses an engineering problem when implementing a time-agnostic imaging solution in the on-board hardware of the sky camera. Therefore, it is   important to develop a sky/cloud segmentation method that achieves competitive performance for both daytime- and nighttime- sky/cloud images. 
We attempt to address this gap in the literature by proposing a cloud image segmentation framework for nychthemeron (i.e.\ the 24-hour timespan that includes a night and a day).

In this paper, we propose a deep-learning architecture called CloudSegNet for efficient cloud segmentation. It is essentially an encoder-decoder architecture -- consisting of convolution-, deconvolution- and pooling layers. It provides a pixel-based semantic segmentation of sky/cloud images, either in the form of a probability map, or as pixel-level binary labels, which identify  each point as either \emph{sky} or \emph{cloud}. 

The main contributions of this paper are the following: %(a) 
\begin{itemize}
\item A light-weight deep-learning architecture for efficient cloud segmentation; and %(b) 
\item A common framework for both daytime and nighttime sky/cloud images, outperforming current methods on either image type.
\end{itemize}

\section{Architecture}

\begin{figure*}[htb]
\begin{center}
\includegraphics[width=0.9\textwidth]{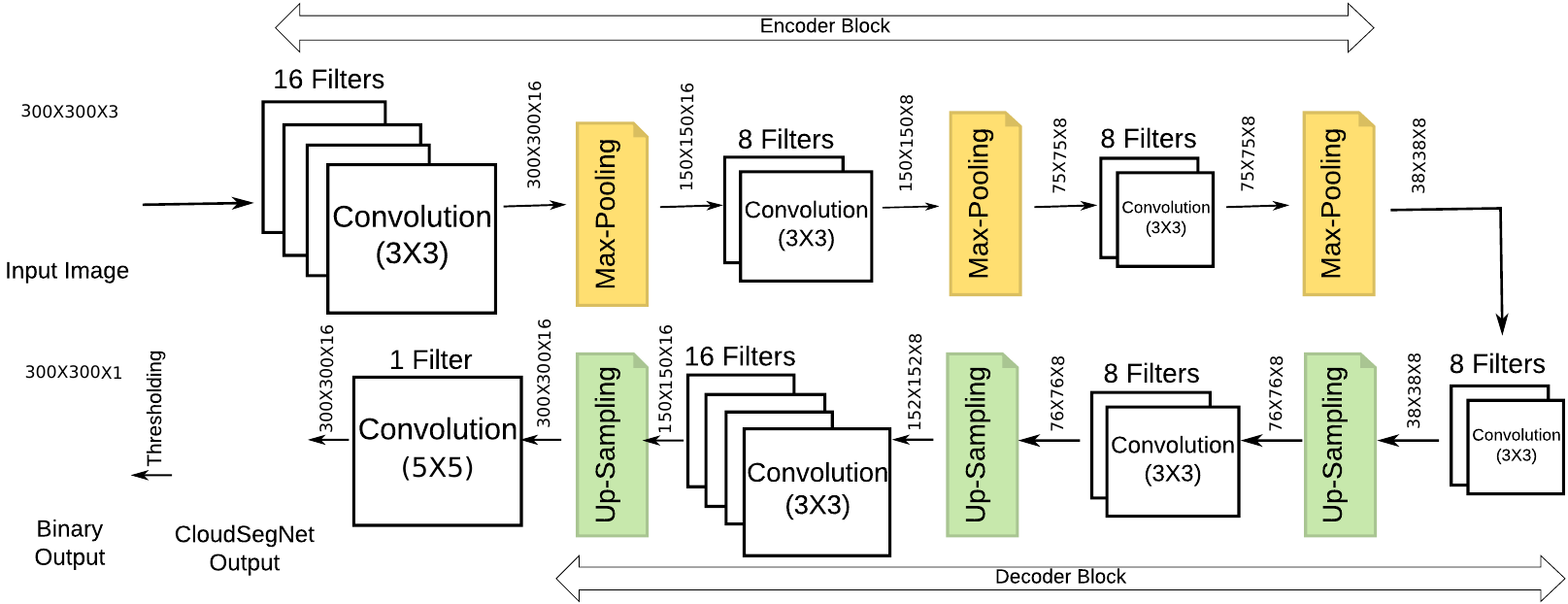}
\caption{Encoder-decoder architecture of the CloudSegNet model. The tensor dimensions at the output of each block are specified. The CloudSegNet output is a  probability mask, which is binarized via thresholding in the final step. 
\label{fig:cloud-model}}
\end{center}
%\vspace{-0.5cm}
\end{figure*}

Our CloudSegNet architecture consists of an encoder-decoder combination, as shown in Fig.~\ref{fig:cloud-model}. It is similar to the convolution layers in VGG16 \cite{simonyan2014very} and SegNet \cite{badrinarayanan2017segNet} networks. The motivation for using such an encoder-decoder architecture is that it helps in aggregating discriminatory image features from different levels and generating a semantic segmentation mask at the same resolution as the input image. Moreover, it contains significantly fewer trainable parameters as compared to other deep neural network architectures. The CloudSegNet architecture does not employ any fully connected layers. Also, it does not use any skip architecture combining deep information with shallow information of the network, as done in \cite{long2015fully}. Instead, we propose a minimalist version of the encoder-decoder architecture for cloud segmentation.

\subsection{Encoder}

The encoder block of CloudSegNet consists of three convolution layers and three max-pooling layers. We feed an RGB image of a fixed input size ($300\times300$ pixels) into the CloudSegNet model. In case the input RGB images are of a different resolution, the images need to be resized to $300\times300$. 
The images are generally normalized using mean subtraction beforehand. We know from \cite{simonyan2014very,he2016deep} that the lower convolution layers compute the primitive image cues viz.\ color, texture. The upper layers compute more complex features using these lower layer features. 
In the encoder block, we use $16$ filters of dimension $3\times3$ for the first convolution layers and $8$ filters of dimension $3\times3$ for the second and third convolution layers. 
The resulting encoder output is of dimension $38\times38\times8$.

\subsection{Decoder}
The decoder block consists of four deconvolution layers and three up-sampling layers. The output of the encoder block (having dimension $38\times38\times8$) is fed into the decoder block. 
In the decoder layers, the first and second deconvolution layers consists of $8$ filters with dimension $3\times3$, and the third deconvolution consists of $16$ filters with dimension $3\times3$.  The final deconvolution layer consists of a single filter with dimension $5\times5$.

\subsection{Implementation}
The output of CloudSegNet is a probability mask, assigning a probabilistic value to each pixel, which represents its likelihood of belonging to the cloud category. We then convert the probability mask into a binary map using a simple thresholding process. The labeling threshold is determined from the Receiver Operating Curve (ROC) of CloudSegNet. 

We implement the CloudSegNet architecture in the TensorFlow framework.\footnote{~The code of all simulations in this paper is available online at \url{https://github.com/Soumyabrata/CloudSegNet}.}  We train the CloudSegNet model from scratch with a dataset of $1128$ images captured by a ground-based sky camera. Our model is trained over $10000$ epochs, using Adadelta optimizer. We use a batch size of $16$, a learning rate of $1.0$, using binary cross entropy loss function. We choose the CloudSegNet model with the lowest validation loss.

\section{Experiments \& Results}

\subsection{Dataset}
We combine images from two publicly available sky/cloud image segmentation datasets -- SWIMSEG (Singapore Whole Sky IMaging SEGmentation dataset) \cite{dev2017color} and SWINSEG (Singapore Whole sky Nighttime Imaging SEGmentation Database) \cite{dev2017nighttime} -- to create a composite dataset of nychthemeron images. These images are undistorted from the original fisheye lens obtained from the sky camera using the camera calibration function. More details about this calibration process can be found in \cite{dev2014wahrsis}.

The SWIMSEG dataset consists of $1013$ daytime sky/cloud images, along with manually annotated ground-truth maps. The SWINSEG dataset consists of $115$ diverse nighttime sky/cloud images, along with the corresponding ground-truth maps. All these images were captured on the rooftop of a building on the campus of Nanyang Technological University (NTU), Singapore by a high-resolution sky camera \cite{dev2014wahrsis}.

Note that the composite dataset is intrinsically unbalanced, as there are more daytime than nighttime images. More discussion on this aspect can be found later in Section~\ref{sec:discuss}.
In order to reduce the impact of imbalance nature of the two datasets, we perform image augmentation techniques for both SWIMSEG- and SWINSEG-datasets. Such augmentation techniques also help the users to train the neural network with varying types of sky/cloud images. We increase the number of images by an additional $5$X order of magnitude. We refer to this augmented composite dataset as Singapore Whole sky Nychthemeron Imaging SEGmentation Database (SWINySeg)~\footnote{The SWINySeg dataset is available for download at \url{http://vintage.winklerbros.net/swinyseg.html}}. 

We use several image-based operations on the original images for the task of image augmentation. We include five types of operations -- rotate, shear, flip, shift, and zoom. Each augmented image is generated as a result of these five distinct operations (applied in sequence), to an original image. All the operations are performed with a random degree of magnitude. We use random magnitude of rotation within a range of $\pm20$ degrees and a resolution of $0.5$ degrees. We use a random magnitude of shear within the range of $\pm0.5$, and a resolution of $0.5$. The shifting operation is also performed within a ratio of $\pm0.1$ in both width and height directions, and a resolution of $0.05$. Finally, the zoom operation is performed within a range $[0.8,1]$ and resolution of $0.05$. We do not perform any transformation in the color, hue or saturation of the sky/cloud image. This is because color is a discriminatory feature for clouds \cite{dev2014systematic}, and altering it would  adversely impact the segmentation performance.

The total number of augmented daytime- and nighttime- images are $5065$ and $575$ respectively. The total number of images in the extended composite \emph{SWINySeg} dataset (inclusive of original and augmented) images is $6768$. Post the augmentation process, we visually inspect all the generated sky/cloud images to make sure that the augmented images look realistic and natural. As all the parameters in the image augmentation process are changed in moderation, our augmented images closely resemble actual sky/cloud images.

\subsection{Loss Trend of CloudSegNet}

\begin{figure}[htb]
\centering
\includegraphics[width=0.4\textwidth]{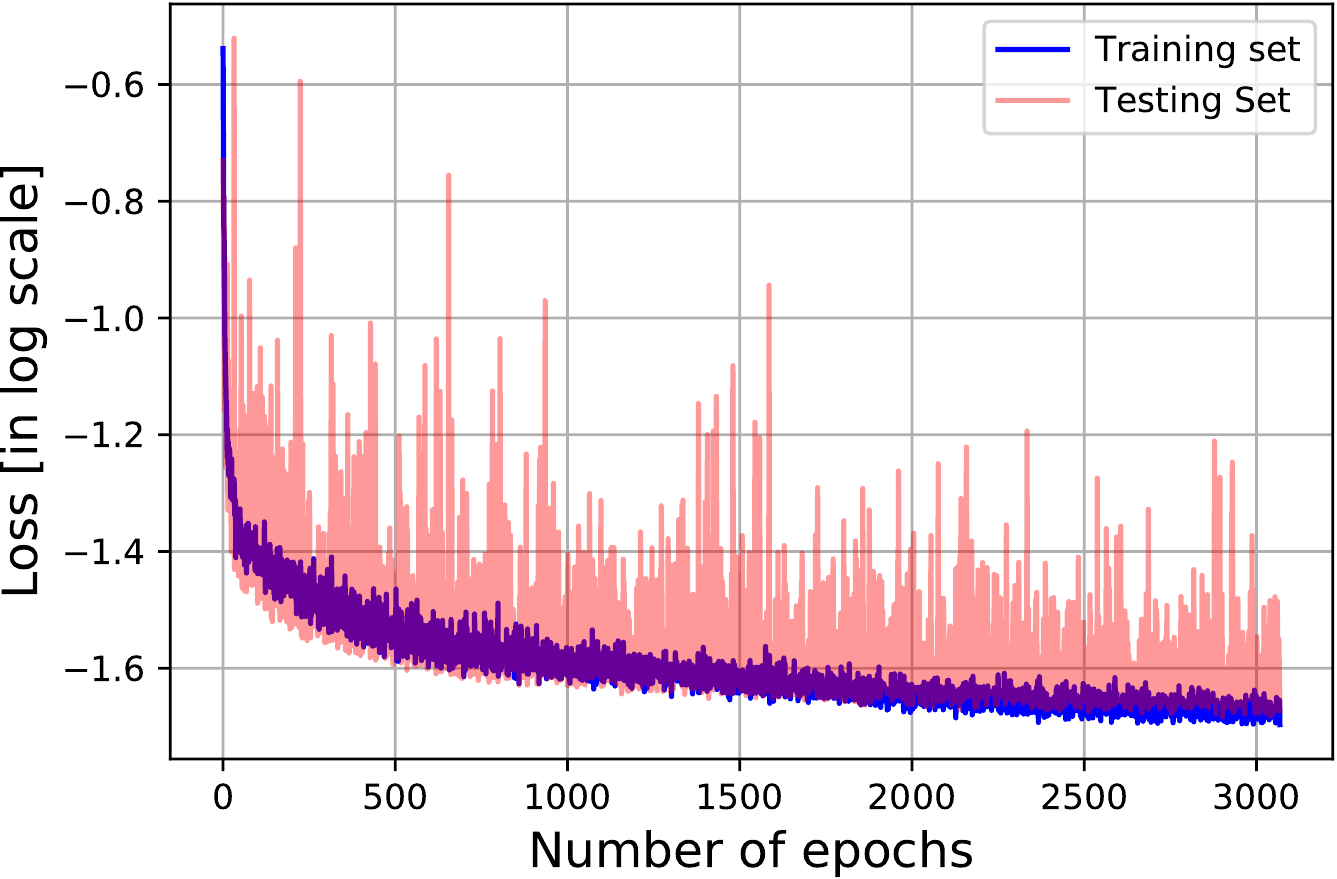}
\caption{Training and testing loss of CloudSegNet model over 3000 epochs.}
\label{fig:loss-trend}
%\vspace{-0.4cm}
\end{figure}

We use this augmented composite SWINySeg dataset of $6768$ images to train the CloudSegNet model. We perform a random sampling on this composite dataset, and divide the training and testing sets in the ratio of $80:20\%$ respectively. The trend of binary cross-entropy loss for training and testing sets are shown in Fig.~\ref{fig:loss-trend}. We observe that the loss saturates after a few thousand iterations, and the model exhibits comparable loss performance for both training  and testing sets.  We choose the CloudSegNet model with the lowest validation loss for our subsequent experiments. 

\subsection{Receiver Operating Curve (ROC)}

\begin{table*}[tb]
\normalsize
\centering
\begin{tabular}{c||r|c|c|c|c}
  \textbf{Image Type}  & \textbf{Method} & \textbf{Precision} & \textbf{Recall} & \textbf{F-score}  & \textbf{Error Rate}  \\
  \hline
  Daytime & Li et al. \cite{li2011a} & 0.81 & \textbf{0.97} & 0.86 & 0.12 \\
  (SWIMSEG)		  & Long et al. \cite{long2006retrieving} & 0.89 & 0.82 & 0.81 & 0.14 \\
          & Souza et al. \cite{souza2006a} & \textbf{0.99} & 0.53 & 0.63 & 0.18 \\
          & Dev et al.\ 2014 \cite{dev2014systematic} & 0.89 & 0.92 & \textbf{0.89} & 0.09 \\
          & Mantelli-Neto et al. \cite{mantelli2010the} & 0.88 & 0.76 & 0.76 & 0.17 \\
          & FCN \cite{long2015fully} & 0.55 & 0.42 & 0.45 & 0.48 \\
          & PSPNet \cite{zhao2017pyramid} & 0.62 & 0.41 & 0.48 & 0.42 \\
          & CloudSegNet & 0.92 & 0.88 & \textbf{0.89} & \textbf{0.07} \\
  \hline          
  Nighttime & Gacal et al. \cite{gacal2016ground} & 0.48 & \textbf{0.99} & 0.62 & 0.50 \\
  (SWINSEG)		    & Yang et al.\ 2009 \cite{yang2009image} & \textbf{0.98} & 0.65 & 0.76 & 0.16 \\
            & Yang et al.\ 2010 \cite{yang2010automatic} & 0.73 & 0.33 & 0.41 & 0.37 \\
            & Dev et al.\ 2017 \cite{dev2017nighttime} & 0.94 & 0.74 & 0.82 & 0.13 \\
            & FCN \cite{long2015fully} & 0.58 & 0.41 & 0.47 & 0.44 \\
            & PSPNet \cite{zhao2017pyramid} & 0.46 & 0.53 & 0.49 & 0.41 \\
            & CloudSegNet & 0.88 & 0.91 & \textbf{0.89} & \textbf{0.08} \\
  \hline
  Day+Night & FCN \cite{long2015fully} & 0.57 & 0.42 & 0.49 & 0.42 \\
          & PSPNet \cite{zhao2017pyramid} & 0.59 & 0.49 & 0.55 & 0.58 \\
          & CloudSegNet & \textbf{0.92} & \textbf{0.87} & \textbf{0.89} & \textbf{0.08} \\
\end{tabular}
\caption{Comparison of CloudSegNet with other cloud detection methods, categorized based on image type. The best performances for each image type are highlighted in bold.}
\label{table:result-table}
%\vspace{-0.6cm}
\end{table*}

\begin{figure}[htb]
\begin{center}
\includegraphics[width=0.4\textwidth]{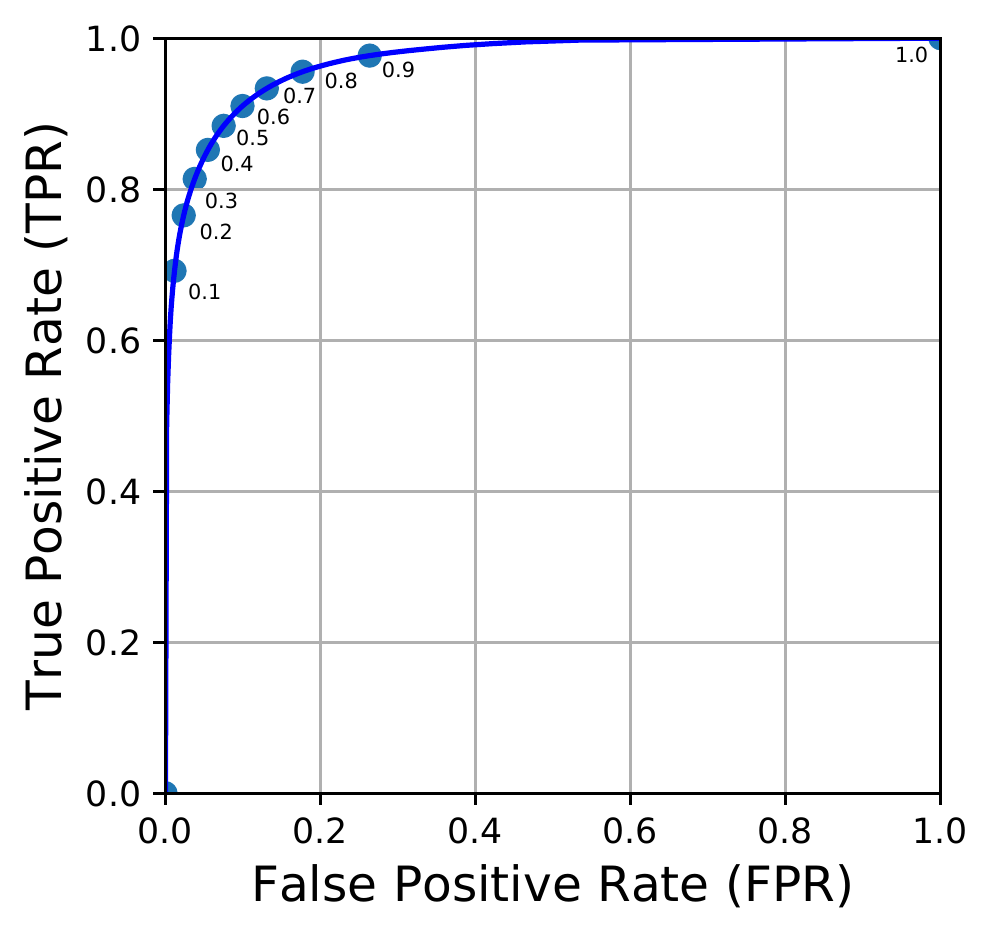}
\caption{ROC curve of our proposed algorithm for varying thresholds in [$0$,$1$]. The various thresholds in steps of $0.1$ are marked along the curve.  
\label{fig:roc-plot}}
\end{center}
%\vspace{-0.4cm}
\end{figure}

As discussed above, the output of CloudSegNet is a  probability mask, wherein each pixel indicates the degree of belongingness to the \emph{cloud} category. Since the ground-truth maps are binary in nature, it is necessary to convert the probabilistic output into binary maps as well. We employ a Receiver Operating Curve (ROC) technique to understand the impact of the threshold on the performance. We vary the threshold from $0$ to $1$ in steps of $0.01$, and record the False Positive Rate (FPR) and True Positive Rate (TPR) of cloud detection. Figure~\ref{fig:roc-plot} shows the resulting ROC curve. The area under the ROC curve (AUC) is $0.97$, indicating the competitive performance of CloudSegNet.

The ROC curve provides an opportunity to choose a threshold, based on the trade-off between FPR and TPR. In our experiments, we choose the threshold of $0.5$ to convert the probabilistic map into a binary sky/cloud image, which is very close to equal true and false positive rates. Of course, this threshold can be further adjusted by the user, based on the specific requirements for TPR and/or FPR. 

\begin{figure}[htb]
\begin{center}
\includegraphics[height=0.11\textwidth]{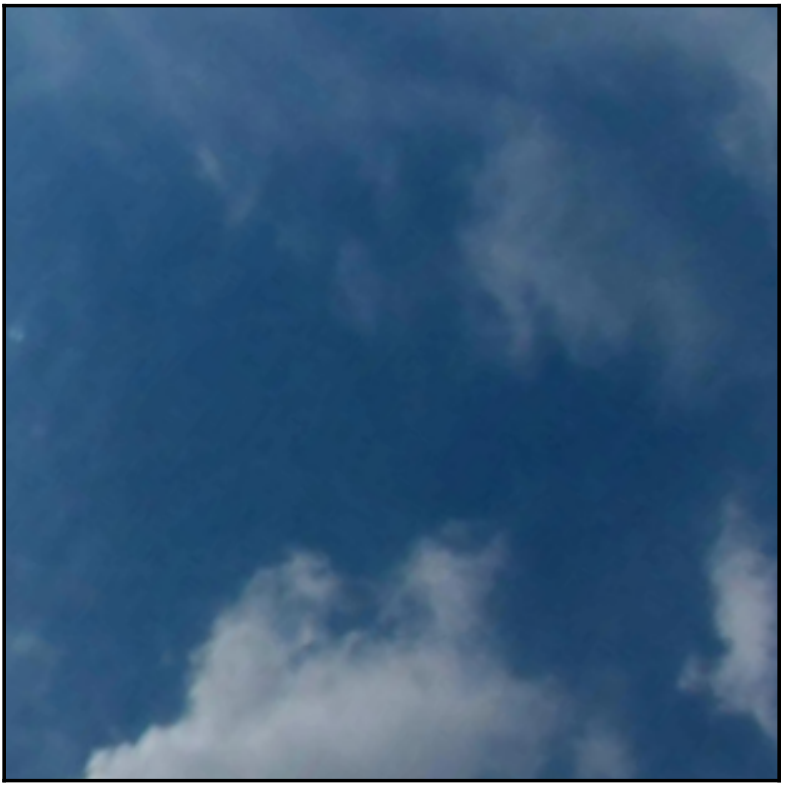}
\includegraphics[height=0.11\textwidth]{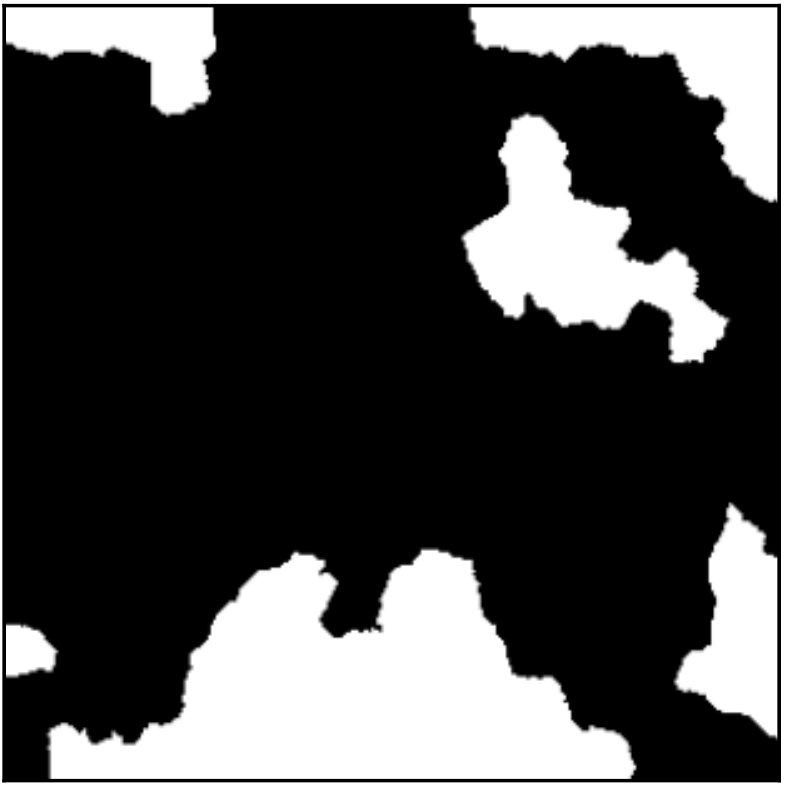}
\includegraphics[height=0.11\textwidth]{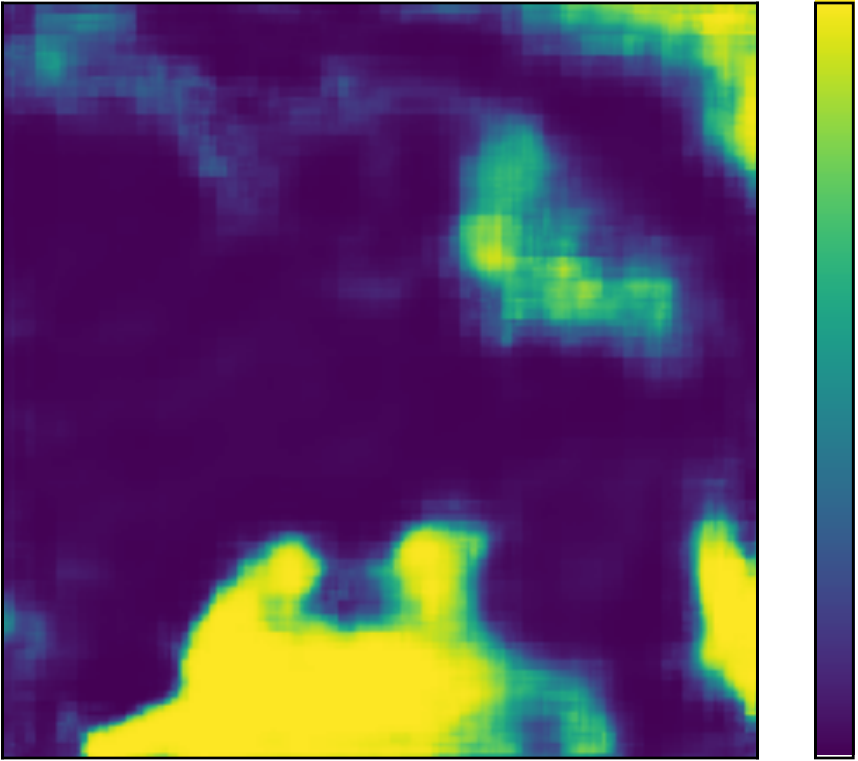}
\includegraphics[height=0.11\textwidth]{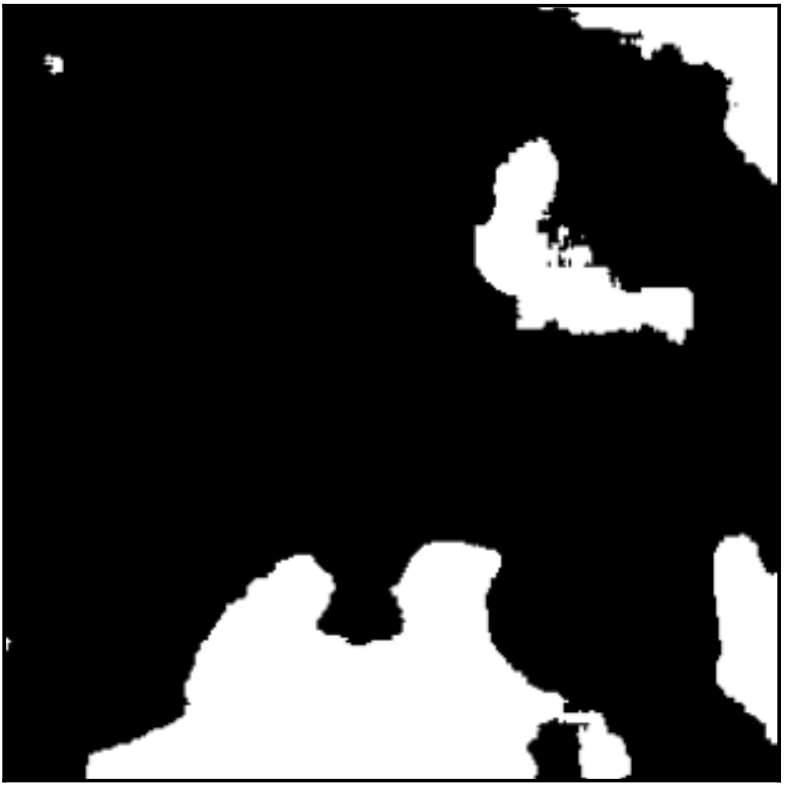}\\
\includegraphics[height=0.11\textwidth]{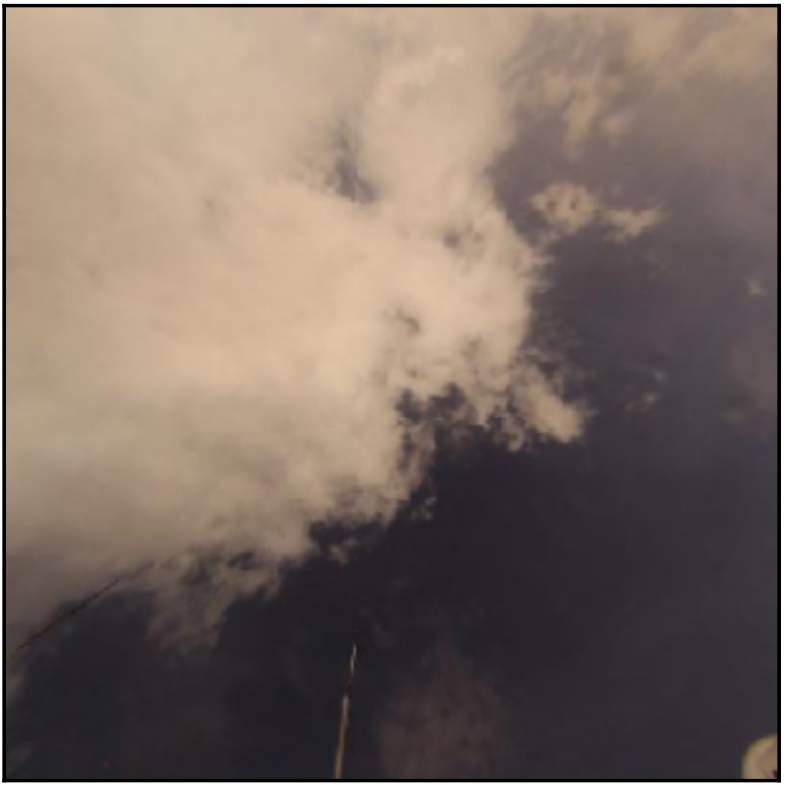}
\includegraphics[height=0.11\textwidth]{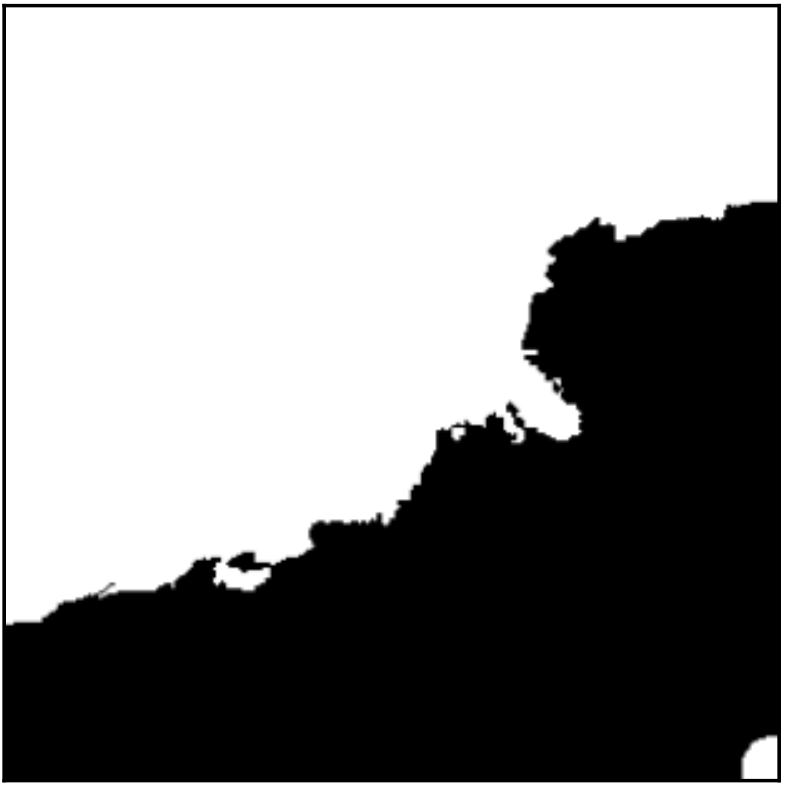}
\includegraphics[height=0.11\textwidth]{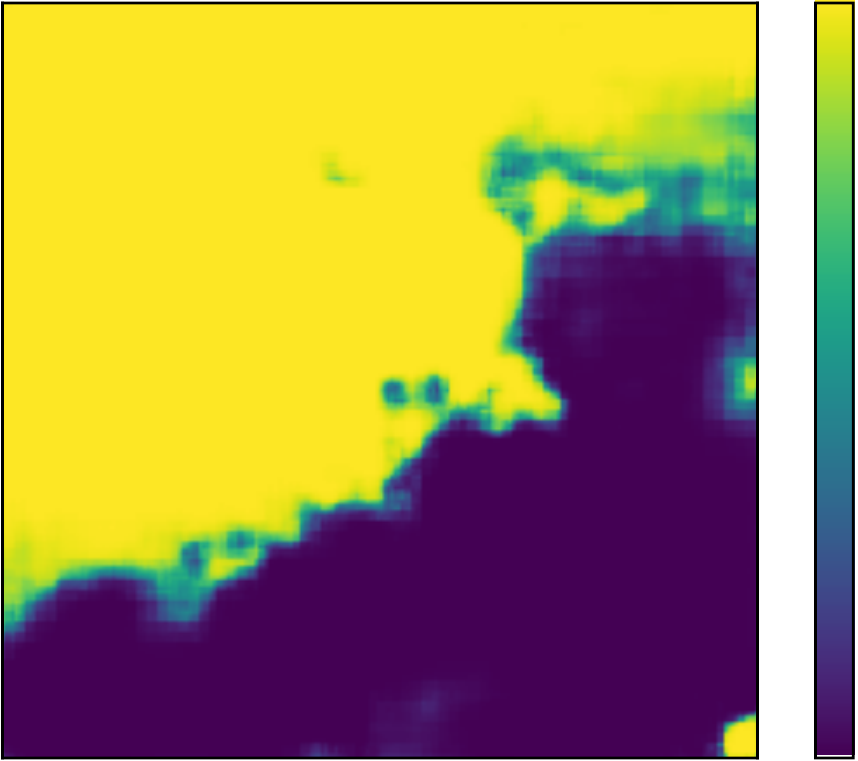}
\includegraphics[height=0.11\textwidth]{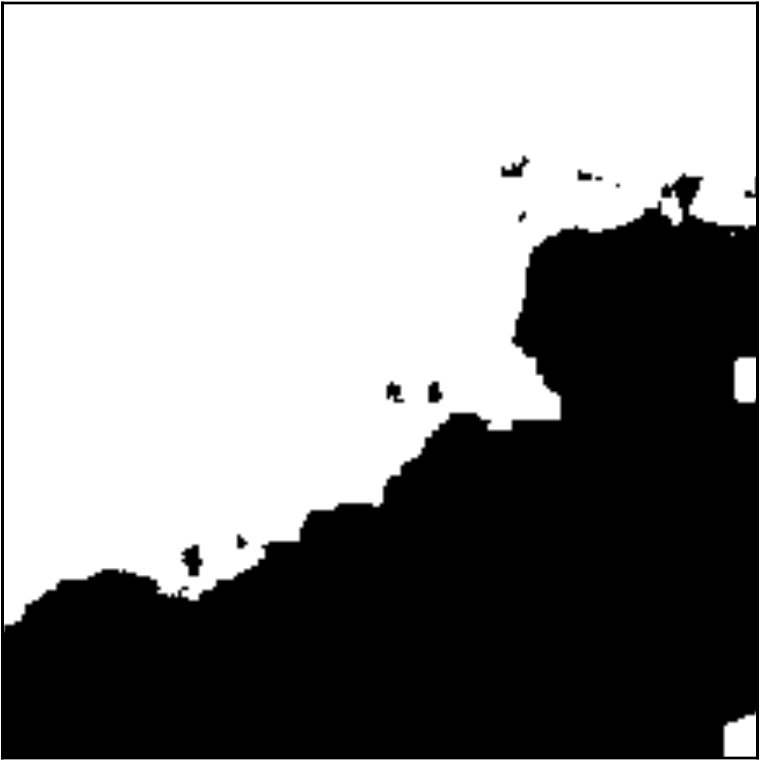}\\
\makebox[0.11\textwidth][c]{\scriptsize Input Image}
\makebox[0.11\textwidth][c]{\scriptsize Ground Truth}
\makebox[0.11\textwidth][c]{\scriptsize Probabilistic}
\makebox[0.11\textwidth][c]{\scriptsize Binary Map}
\caption{Examples of cloud/sky segmentation for daytime (top row) and nighttime (bottom row) images using CloudSegNet.  The probability map shows the degree of \emph{belongingness} to cloud. We threshold this map into a binary sky/cloud image.
\label{fig:visual}}
\end{center}
%\vspace{-1cm}
\end{figure}

Figure~\ref{fig:visual} shows some sample outputs of our proposed approach.  Visual inspection of additional images from our SWINySeg dataset confirms that CloudSegNet can successfully identify cloud pixels from nychthemeron images. 

\subsection{Benchmarking}

We benchmark the performance of CloudSegNet with current state-of-the-art cloud detection algorithms. In the literature, there are no algorithms or frameworks that can \emph{simultaneously} work for both daytime and nighttime; they are intended for either day-only or night-only sky/cloud images. Li et al.\ \cite{li2011a} use a set of fixed and adaptive thresholds in the ratio color channel of red and blue color channels, to generate the binary map. Souza et al.\ \cite{souza2006a} use the saturation component in Intensity-Hue-Saturation (IHS) color model to detect clouds in the captured image. Dev et al.\ \cite{dev2014systematic} use a clustering technique on the ratio channel of red and blue channels. Mantelli-Neto et al.\ use the RGB color model for cloud detection \cite{mantelli2010the}. On the other hand, Long et al.\ \cite{long2006retrieving} attempt to model the atmospheric scattering, and provide a threshold for efficient cloud detection. 

Nighttime sky/cloud image segmentation is much less studied. Recently, Gacal et al.\ analyzed  clouds in the Philippines, and proposed a threshold in the gray color channel of the image \cite{gacal2016ground}. Yang et al.\ (2009) \cite{yang2009image} used Otsu thresholding approach on the difference of red and blue  channels to generate the binary image. Yang et al.\ (2010) \cite{yang2010automatic} used efficient adaptive thresholds on sub-images to generate the binary cloud image. Dev et al.\ \cite{dev2017nighttime} used a super-pixel based technique for nighttime cloud segmentation.

We benchmark our CloudSegNet approach with these various daytime- and nighttime- image segmentation approaches. In order to provide a quantitative evaluation of our proposed method, we report Precision, Recall, F-score and Error Rate of cloud detection. Suppose, $TP$, $FP$, $TN$ and $FN$ denote the true positives, false positives, true negatives and false negatives of our binary cloud detection problem. The different metrics are defined as follows: $\mbox{Precision} = \frac{TP}{TP+FP}$, $\mbox{Recall} = \frac{TP}{TP+FN}$, $\mbox{F-Score} = (2\times\mbox{Precision}\times\mbox{Recall})/(\mbox{Precision}+\mbox{Recall})$ and $\mbox{Error Rate} = \frac{FP+FN}{TP+FP+TN+FN}$.

These metrics are calculated on the testing images of our SWINySeg dataset. Since the existing algorithms are tested on day-only- or night-only- sky/cloud images, we measure these metrics separately, based on the image type. However, it is important to note  that CloudSegNet is trained only once on the composite dataset (consisting of both daytime- and nighttime- images), even though the benchmarking is done separately. 

We present our benchmarking results in Table~\ref{table:result-table}. For daytime images, Souza et al.\ achieves a very high precision, while Li et al.\ has a high recall. However, CloudSegNet achieves the best overall performance with highest F-score and lowest Error Rate. Similarly, for nighttime images, Yang et al.\ (2009) obtains the highest precision, while Gacal et al.\ exhibits the highest  recall. Our proposed CloudSegNet method again achieves the best performance with respect to F-score and Error Rate, despite the imbalance in the composite dataset. CloudSegNet is able to maintain this competitive performance on the SWINySeg dataset of daytime- and nighttime- testing images. 

\subsection{Discussion}
\label{sec:discuss}
Our composite SWINySeg dataset consisting of daytime- and nighttime- sky/cloud images is unbalanced in nature. It contains $5065$ daytime images, and only $575$ nighttime images, along with their corresponding binary ground-truth labels. In this section, we analyze the impact of this unbalanced nature on the classification metric. We perform random down-sampling, such that the ratio of number of nighttime- and the number of daytime- images is $50:50\%$. In this experiment, we consider all the $575$ nighttime images of the SWINSEG dataset. In addition to it, we randomly select $575$ daytime images of the SWIMSEG dataset to create a balanced dataset of sky/cloud images. This balanced dataset of nychthemeron sky/cloud images contains an equal number of daytime- and nighttime- images.

We train CloudSegNet model from scratch on this balanced dataset, containing $575$ daytime and $575$ nighttime images. In each experiment, we consider all the nighttime images, and perform a random selection of $575$ daytime images. The CloudSegNet model is then subsequently trained on a random $80\%$ subset of the balanced dataset, while tested on the remaining $20\%$ of the images. After training, we evaluate the performance of the trained model on daytime, nighttime, and nychthemeron sky/cloud images. We perform $10$ such experiments with uniform downsampling strategy to remove any sample bias. We use the same hyperparameter values to train on this balanced dataset for all experiments. 

Table~\ref{table:balanced-result} summarises the performance of CloudSegNet in the balanced dataset, with respect to $200$ distinct experiments. We observe that the performance of the CloudSegNet model is competitive in the balanced dataset too -- the F-score value for daytime, nighttime and nychthemeron sky/cloud images are $0.90$, $0.93$, and $0.92$. This indicates that the unbalanced nature of the composite dataset does not have any adverse impact on segmentation accuracy.  Although the CloudSegNet model is trained on the composite dataset of nychthemeron sky/cloud images, we achieve good segmentation accuracy on both daytime- and nighttime- images individually. 

\begin{table}[htb]
\normalsize   
\centering
\begin{tabular}{l||c|c|c|c}
  \textbf{Image Type}  & \textbf{Precision} & \textbf{Recall} & \textbf{F-score}  & \textbf{Error Rate}  \\
  \hline
  Daytime & 0.94 & 0.93 & 0.90 & 0.07 \\     
  Nighttime & 0.94 & 0.94 & 0.93 & 0.06 \\
  Day+Night & 0.94 & 0.93 & 0.92 & 0.07 \\
\end{tabular}
\caption{Performance evaluation of CloudSegNet on the balanced cloud segmentation dataset, categorized based on image type. We report the average value of these evaluation metrics for all the different image types.}
\label{table:balanced-result}
%\vspace{-0.2cm}
\end{table}

\balance 

\section{Conclusions}
\label{sec:conclusion}
We presented the first deep-learning architecture for nychthemeron (day \& night) cloud image segmentation. We showed the competitive performance of our proposed method on a combination of two  publicly available sky/cloud image segmentation datasets. We use image-based augmentation techniques to increase the size of training and testing sets.  Our method does not need a careful selection of color channels for discriminatory input feature -- CloudSegNet learns the most discriminatory color and texture features for efficient cloud segmentation. Our future work includes using such deep neural networks in solving other imaging problems, from ground-based sky/cloud images. We also intend to release a larger dataset of sky/cloud images with manually annotated labels to the remote sensing community.


\begin{thebibliography}{10}

\bibitem{dev2016ground}
S.~Dev, B.~Wen, Y.~H. Lee, and S.~Winkler,
\newblock ``Ground-based image analysis: A tutorial on machine-learning
  techniques and applications,''
\newblock {\em IEEE Geoscience and Remote Sensing Magazine}, vol. 4, no. 2, pp.
  79--93, June 2016.

\bibitem{cheng2018when}
G.~{Cheng}, C.~{Yang}, X.~{Yao}, L.~{Guo}, and J.~{Han},
\newblock ``When deep learning meets metric learning: Remote sensing image
  scene classification via learning discriminative {CNNs},''
\newblock {\em IEEE Transactions on Geoscience and Remote Sensing}, vol. 56,
  no. 5, pp. 2811--2821, May 2018.

\bibitem{li2018deep}
Y.~Li, Y.~Zhang, X.~Huang, and A.~L. Yuille,
\newblock ``Deep networks under scene-level supervision for multi-class
  geospatial object detection from remote sensing images,''
\newblock {\em ISPRS Journal of Photogrammetry and Remote Sensing}, vol. 146,
  pp. 182--196, 2018.

\bibitem{li2011a}
Q.~Li, W.~Lu, and J.~Yang,
\newblock ``A hybrid thresholding algorithm for cloud detection on ground-based
  color images,''
\newblock {\em Journal of Atmospheric and Oceanic Technology}, vol. 28, no. 10,
  pp. 1286--1296, Oct. 2011.

\bibitem{souza2006a}
M.~P. Souza-Echer, E.~B. Pereira, L.~S. Bins, and M.~A.~R. Andrade,
\newblock ``A simple method for the assessment of the cloud cover state in
  high-latitude regions by a ground-based digital camera,''
\newblock {\em Journal of Atmospheric and Oceanic Technology}, vol. 23, no. 3,
  pp. 437--447, March 2006.

\bibitem{dev2017rough}
S.~{Dev}, F.~M. {Savoy}, Y.~H. {Lee}, and S.~{Winkler},
\newblock ``Rough-set-based color channel selection,''
\newblock {\em IEEE Geoscience and Remote Sensing Letters}, vol. 14, no. 1, pp.
  52--56, Jan 2017.

\bibitem{long2006retrieving}
C.~N. Long, J.~M. Sabburg, J.~Calb\'{o}, and D.~Pages,
\newblock ``Retrieving cloud characteristics from ground-based daytime color
  all-sky images,''
\newblock {\em Journal of Atmospheric and Oceanic Technology}, vol. 23, no. 5,
  pp. 633--652, 2006.

\bibitem{dev2014systematic}
S.~Dev, Y.~H. Lee, and S.~Winkler,
\newblock ``Systematic study of color spaces and components for the
  segmentation of sky/cloud images,''
\newblock in {\em Proc. International Conference on Image Processing (ICIP)},
  2014, pp. 5102--5106.

\bibitem{yang2009image}
Q.~Yang, L.~Tang, W.~Dong, and Y.~Sun,
\newblock ``Image edge detecting based on gap statistic model and relative
  entropy,''
\newblock in {\em Proc. 4th International Conference on Fuzzy Systems and
  Knowledge Discovery}, 2009.

\bibitem{yang2010automatic}
J.~Yang, W.~T. Lv, Y.~Ma, W.~Yao, and Q.~Y. Li,
\newblock ``An automatic ground-based cloud detection method based on local
  threshold interpolation,''
\newblock {\em Acta Meteorologica Sinica}, vol. 68, no. 6, pp. 1007--1017,
  2010.

\bibitem{simonyan2014very}
K.~Simonyan and A.~Zisserman,
\newblock ``Very deep convolutional networks for large-scale image
  recognition,''
\newblock {\em CoRR}, vol. abs/1409.1556, 2014.

\bibitem{badrinarayanan2017segNet}
V.~Badrinarayanan, A.~Kendall, and R.~Cipolla,
\newblock ``{SegNet}: A deep convolutional encoder-decoder architecture for
  image segmentation,''
\newblock {\em IEEE Transactions on Pattern Analysis and Machine Intelligence},
  vol. 39, no. 12, pp. 2481--2495, 2017.

\bibitem{long2015fully}
J.~Long, E.~Shelhamer, and T.~Darrell,
\newblock ``Fully convolutional networks for semantic segmentation,''
\newblock in {\em Proc. IEEE Conference on Computer Vision and Pattern
  Recognition (CVPR)}, 2015, pp. 3431--3440.

\bibitem{he2016deep}
K.~He, X.~Zhang, S.~Ren, and J.~Sun,
\newblock ``Deep residual learning for image recognition,''
\newblock in {\em Proc. IEEE Conference on Computer Vision and Pattern
  Recognition (CVPR)}, 2016, pp. 770--778.

\bibitem{dev2017color}
S.~Dev, Y.~H. Lee, and S.~Winkler,
\newblock ``Color-based segmentation of sky/cloud images from ground-based
  cameras,''
\newblock {\em IEEE Journal of Selected Topics in Applied Earth Observations
  and Remote Sensing}, vol. 10, no. 1, pp. 231--242, 2017.

\bibitem{dev2017nighttime}
S.~Dev, F.~M. Savoy, Y.~H. Lee, and S.~Winkler,
\newblock ``Nighttime sky/cloud image segmentation,''
\newblock in {\em Proc. International Conference on Image Processing (ICIP)},
  2017.

\bibitem{dev2014wahrsis}
S.~Dev, F.~M. Savoy, Y.~H. Lee, and S.~Winkler,
\newblock ``{WAHRSIS}: A low-cost, high-resolution whole sky imager with
  near-infrared capabilities,''
\newblock in {\em Proc. IS\&T/SPIE Infrared Imaging Systems}, 2014.

\bibitem{mantelli2010the}
S.~L. Mantelli~Neto, A.~von Wangenheim, E.~B. Pereira, and E.~Comunello,
\newblock ``The use of {Euclidean} geometric distance on {RGB} color space for
  the classification of sky and cloud patterns,''
\newblock {\em Journal of Atmospheric and Oceanic Technology}, vol. 27, no. 9,
  pp. 1504--1517, 2010.

\bibitem{zhao2017pyramid}
H.~Zhao, J.~Shi, X.~Qi, X.~Wang, and J.~Jia,
\newblock ``Pyramid scene parsing network,''
\newblock in {\em Proc. IEEE Conf. on Computer Vision and Pattern Recognition
  (CVPR)}, 2017, pp. 2881--2890.

\bibitem{gacal2016ground}
G.~F.~B. Gacal, C.~Antioquia, and N.~Lagrosas,
\newblock ``Ground-based detection of nighttime clouds above {Manila}
  observatory (14.64 {N}, 121.07 {E}) using a digital camera,''
\newblock {\em Applied Optics}, vol. 55, no. 22, pp. 6040--6045, 2016.

\end{thebibliography}
\end{document}